# CT-based Anomaly Detection of Liver Tumors Using Generative Diffusion Prior

Yongyi Shi, Chuang Niu, Amber L. Simpson, Bruno De Man, Richard Do, Ge Wang

*Abstract*— **CT is a main modality for imaging liver diseases, valuable in detecting and localizing liver tumors. Traditional anomaly detection methods analyze reconstructed images to identify pathological structures. However, these methods may produce suboptimal results, overlooking subtle differences among various tissue types. To address this challenge, here we employ generative diffusion prior to inpaint the liver as the reference facilitating anomaly detection. Specifically, we use an adaptive threshold to extract a mask of abnormal regions, which are then inpainted using a diffusion prior to calculating an anomaly score based on the discrepancy between the original CT image and the inpainted counterpart. Our methodology has been tested on two liver CT datasets, demonstrating a significant improvement in detection accuracy, with a 7.9% boost in the area under the curve (AUC) compared to the state-of-the-art. This performance gain underscores the potential of our approach to refine the radiological assessment of liver diseases.**

## I. INTRODUCTION

LIVER cancer is the second leading cause of cancer death among men globally, with a five-year survival rate below 18% [1, 2]. Despite the risk of ionizing radiation exposure, CT is a main diagnostic tool due to its detailed tomographic images, relatively low cost, and high scan speed [3]. In this context, CT image analysis presents a significant challenge due to the complexity and subtleties of human anatomy and pathology, especially the lack of intensity/textural contrast between tissue types, variability of tissue properties, and presence of noise [4]. As a result, this process relies heavily on the radiologists' subjective judgement, potentially compromising diagnostic accuracy and consistency [5]. Additionally, identifying tumors requires radiologists to thoroughly analyze each CT slice, which is tedious and time-consuming [6]. Since decades ago, computer-aided diagnosis (CAD) systems have been under active development [7-8]. These systems automate the detection and localization of tumors, helping radiologists improve the diagnostic performance [9].

In recent years, deep learning-based CAD has made great strides [10]. Inspired by the U-Net [11], a variety of its variants [12], including U-Net++, Swin U-Net, and nnU-Net [13-15], emerged to address diverse challenges in medical image analysis. Despite the resultant promising results, even achieving accuracies comparable to clinical experts [16], these models predominantly rely on supervised learning, necessitating extensive, high-quality annotations for end-to-end training. However, the manual creation of pixel-wise annotations is time-intensive, expert-dependent, and highly costly, resulting in rather limited availability of labeled data. Moreover, radiologist-generated labels are prone to errors and biases, undermining the outcomes of CAD systems [17].

Interestingly, collection of normal medical CT images from healthy subjects is common and comes with labels naturally (negative radiology reports). Logically, anything significantly different from the normal CT images should be considered as abnormal. Therefore, this is an unsupervised way for anomaly detection without annotating tumors [18].

Anomaly detection involves identifying features that deviate from those found in a normal distribution. The anomaly detection process is analogous to the learning process for radiologists to recognize healthy anatomical structures first and then detect abnormalities, even without specific prior knowledge of their pathological attributes. Thus, the hypothesis is that by capturing the distribution of healthy anatomical structures through training a deep generative model, anomalies can be identified as outliers relative to this normative distribution [19-20].

Tomographic methods are a dominant strategy for medical anomaly detection. Numerous studies reported the use of vanilla autoencoder (AE) and its variants [21-22], including variational AE (VAE) [23], perceptual AE [24], adversarial AE [25], memory-augmented AE [26], and heterogeneous AE [27]. Although these methods allow a stable training process, the quality of the reconstructed images is often suboptimal. To mitigate this issue, generative adversarial networks (GANs) were used to replace AE and produce high-quality images [28-30]. Along this direction, AnoVAE-GAN [31] and GANomaly [32] introduced adversarial training to AE-based models and

This work was supported in part by the National Institute of Biomedical Imaging and Bioengineering/National Institutes of Health under Grant R01 CA233888 (Corresponding authors: Ge Wang).

Yongyi Shi, Chuang Niu and Ge Wang are with Biomedical Imaging Center, Department of Biomedical Engineering, School of Engineering, Rensselaer Polytechnic Institute, Troy, NY 12180 USA (email: shiy11@rpi.edu; niuc@rpi.edu; wangg6@rpi.edu).

Amber L. Simpson is with Biomedical Computing and Informatics, Queen's University, ON K7L 3N6, Canada (email: amber.simpson@queensu.ca).

Bruno De Man is with GE HealthCare Technology & Innovation Center, Niskayuna, NY 12309 USA (email: deman@gehealthcare.com).

Richard Do is with Department of Radiology, Memorial Sloan Kettering Cancer Center, New York, NY 10065 USA (email: dok@mskcc.org).

enhanced the image quality. Furthermore, denoising AE incorporates noise into the training data, producing high-quality images [33].

Despite the promising results achieved by current models, they suffer from the issue that abnormal regions may also be reconstructed, making them less distinguishable. To address this issue, researchers proposed erasing partial information from the input image and subsequently predicting the erased content, which is an image inpainting problem [34]. Inpainting helps anomaly detection by synthesizing alternative contents in the missing parts of an image with semantically meaningful normal features to make a realistic-looking health image. For instance, Sogancioglu et al. investigated three inpainting models for context encoding, semantic image inpainting, and contextual attention respectively to perform chest x-ray anomaly detection [35]. Nguyen et al. trained a deep convolutional neural network on normal MRI images to reconstruct healthy brain regions and identify anomalous regions in terms of a high reconstruction loss [36]. Swiecicki et al. trained an inpainting GAN on normal digital breast tomosynthetic images and completely removed parts of interest during inference [37]. Astaraki proposed an auto-inpainting pipeline to automatically detect tumors, replacing their appearance with the learned healthy anatomy [38].

Recently, diffusion models were reported to produce high-quality images while offering distribution coverage, stable training, and easy scalability [39, 40]. Impressively, the denoising diffusion probabilistic model (DDPM) method was employed for image inpainting [41, 42]. However, these approaches were trained with a certain mask distribution, which can lead to poor generalization to novel mask types. Considering the variable shape of the liver across humans, an inpainting approach needs to perform mask-adaptive training. Although several diffusion model-based methods were proposed for image inpainting with arbitrary masks [43-45], these methods have not been evaluated in the context of medical imaging. Directly applying these approaches to inpaint the whole liver may result in a mismatch between the inpainted region and the original liver image.

In this paper, we propose a generative diffusion prior inpainting-based (GDPI) anomaly detection method for CT liver tumor detection. Specifically, we train a diffusion model on healthy CT images [46]. During inference, we adaptively locate, remove and inpaint potential anomalous regions. Subsequently, we utilize the inpainting results to obtain the anomaly score. Finally, post-processing is applied to the anomaly score to suppress noise and obtain the final result.

We summarize our main contribution as follows:
1) We design an adaptive threshold approach to adaptively locate potential anomalous regions, allowing the inpainting network to inpaint the image more precisely.
2) We introduce a unique inpainting method for CT liver anomaly detection. While semantically inpainting the masked areas, our deep network also recovers textures effectively, mitigating the limitations of autoencoders in anomaly detection.
3) Our method significantly improves the performance of inpainting-based anomaly detection on two CT liver datasets against the competing methods.

The remainder of the paper is organized as follows. Section II describes the proposed GDPI method. Section III presents our experimental design. Section IV reports the experimental outcomes. Section V discusses the relevant issues, and finally Section VI draws the conclusion.

## II. METHODS

Our proposed GDPI method is illustrated in Fig. 1. There are the four main modules in the pipeline as follows. In the training phase, the DDPM is trained on a dataset consisting only of healthy CT images. In the test phase, an adaptive mask extraction (AME) module is employed to suggest potential abnormal regions. Then, an inpainting module is used to fuse healthy tissues into these regions. Finally, a scoring module assesses the abnormal regions with the anomaly score for radiologist reading. The detailed description of these four modules is in the following sub-sections.

### A. Diffusion Model

We briefly review the formulation of DDPMs as presented by Ho et al. [47]. DDPM begins with a forward process that progressively adds noise to a normal-dose CT image $x_0 \sim q(x_0)$ over $T$ timesteps, according to a variance schedule $\beta_1, \cdots, \beta_T$:

$$q(x_t|x_{t-1}) = \mathcal{N}(x_t; \sqrt{1-\beta_t}x_{t-1}, \beta_t I) \quad (1)$$

$$q(x_{1:T}|x_0) = \prod_{t=1}^{T} q(x_t|x_{t-1}) \quad (2)$$

where $x_1, \cdots, x_T$ are latent variables of the same dimensionality as the sample $x_0 \sim q(x_0)$.

According to the properties of the Gaussian distribution, the sampling result $x_t$ at an arbitrary timestep $t$ can be written in the following closed form:

$$q(x_t|x_0) = \mathcal{N}(x_t; \sqrt{\bar{\alpha}_t}x_0, (1-\bar{\alpha}_t)I) \quad (3)$$

where $\alpha_t = 1 - \beta_t$ and $\bar{\alpha}_t = \prod_{i=1}^{t} \alpha_i$.

After the forward process, $x_T$ follows a standard normal distribution when $T$ is large enough. Thus, if we know the conditional distribution $q(x_{t-1}|x_t)$, we can use the reverse process to get a sample under $q(x_0)$ from $x_T \sim \mathcal{N}(0, I)$. A neural network can be designed to gradually denoise a Gaussian field, which corresponds to learning the reverse process of a fixed Markov Chain of length $T$. The reverse process can be expressed as

$$p_\theta(x_{t-1}|x_t) = \mathcal{N}(x_{t-1}; \mu_\theta(x_t, t), \sigma_t^2 I) \quad (4)$$

$$p_\theta(x_{0:T}) = p(x_T) \prod_{t=1}^{T} p_\theta(x_{t-1}|x_t) \quad (5)$$

where $p(x_T)$ is the density function of $x_T$. In Eq. (5), $\mu_\theta(x_t, t)$ and $\sigma_t^2$ are needed to solve $p_\theta(x_{t-1}|x_t)$. According to the Bayes theorem, the posterior $q(x_{t-1}|x_t, x_0)$ are defined as:

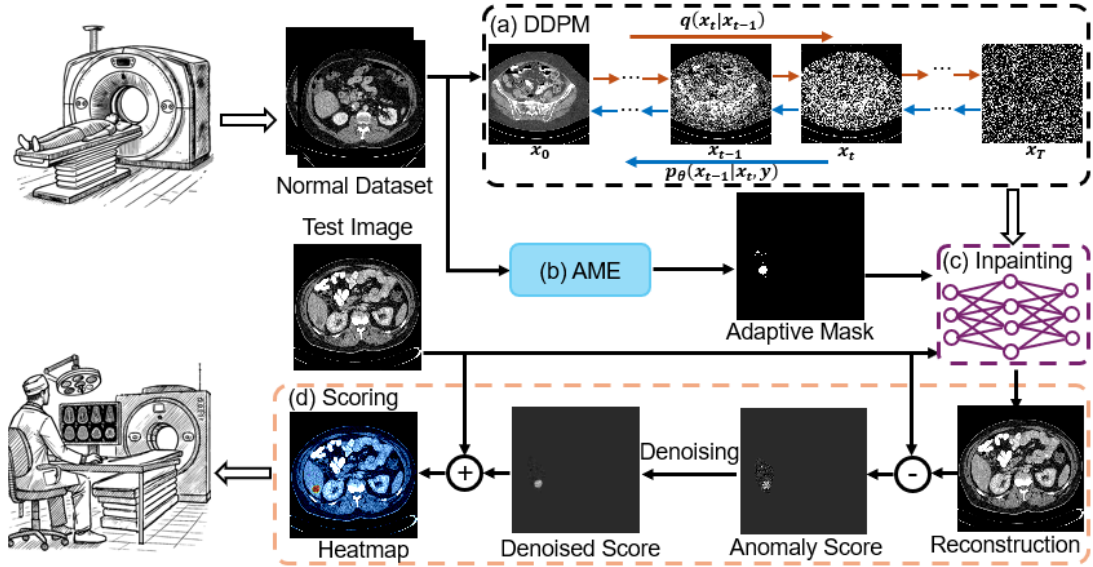

Figure 1. Schematic overview of the proposed GDPI method. The method includes the four key modules: (a) DDPM module: To obtain the generative diffusion prior as a diffusion model trained on normal CT images; (b) AME module: To extract the abnormal region adaptively; (c) Inpainting module: To inpaint the masked regions with healthy tissues; and (d) Scoring module: To highlight abnormal regions for radiologist reading.

$$q(x_{t-1}|x_t, x_0) = \mathcal{N}(x_{t-1}; \tilde{\mu}_t(x_t, x_0), \sigma_t^2 I) \quad (6)$$

where

$$\tilde{\mu}_t(x_t, x_0) = \frac{\sqrt{\alpha_t}(1-\bar{\alpha}_{t-1})}{1-\bar{\alpha}_t}x_t + \frac{\sqrt{\bar{\alpha}_{t-1}}(1-\alpha_t)}{1-\bar{\alpha}_t}x_0 \quad (7)$$

$$\sigma_t^2 = \frac{(1-\bar{\alpha}_{t-1})(1-\alpha_t)}{1-\bar{\alpha}_t} \quad (8)$$

Since $\sigma_t^2$ is a constant, the most natural parameterization of $\mu_\theta(x_t, t)$ is a neural network that predicts $\tilde{\mu}_t(x_t, x_0)$ directly. Alternatively, given that $x_t = \sqrt{\bar{\alpha}_t}x_0 + \sqrt{1-\bar{\alpha}_t}\epsilon, \epsilon \sim \mathcal{N}(0, I)$, the posterior expectation can be expressed as:

$$\tilde{\mu}_t(x_t, x_0) = \tilde{\mu}_t\left(x_t, \frac{1}{\sqrt{\bar{\alpha}_t}}(x_t - \sqrt{1-\bar{\alpha}_t}\epsilon)\right)$$

$$= \frac{1}{\sqrt{\alpha_t}}\left(x_t - \frac{1-\alpha_t}{\sqrt{1-\bar{\alpha}_t}}\epsilon\right) \quad (9)$$

Since $\tilde{\mu}_t(x_t, x_0)$ can be represented by $\epsilon$, we can also use a neural network model $D_\theta$ to predict the noise $\epsilon$. Hence, the corresponding objective can be written as:

$$\mathcal{L} = \mathbb{E}_{x,y}\mathbb{E}_{\epsilon,t}\left[\frac{(1-\alpha_t)^2}{2\sigma_t^2 \alpha(1-\bar{\alpha})}\left\|\epsilon - D_\theta(\sqrt{\bar{\alpha}_t}x_0 + \sqrt{1-\bar{\alpha}_t}\epsilon, t)\right\|_2^2\right] \quad (10)$$

with $t$ uniformly samples $\{1, \cdots, T\}$.

In this study, the latent space is diffused into Gaussian noise after $T = 1000$ steps. A U-Net model $D_\theta$ is trained to predict the noise $\epsilon$ in the latent space. To obtain a high-quality primary content, samples are computed as follows:

$$x_{t-1} = \frac{1}{\sqrt{\alpha_t}}\left(x_t - \frac{1-\alpha_t}{\sqrt{1-\bar{\alpha}_t}}D_\theta(x_t, t)\right) + \sigma_t z \quad (11)$$

where $z \sim \mathcal{N}(0, I)$.

To improve the image quality at a reasonably fast sampling speed, an improved DDPM is used to train the diffusion model on our health liver CT dataset; see [46] for more details.

### B. Adaptive Mask Extraction

To detect liver tumors in a CT image, an effective strategy is to use the previously described generative diffusion prior to inpaint the liver region. However, directly inpainting the entire liver region may result in a mismatch between the inpainted features and the original counterparts. This can cause disagreements in the liver region, compromising the final abnormality detection performance.

To enhance inpainting accuracy, we utilize an AME module to suggest candidate abnormal tumor regions. This allows more healthy liver regions to be retained, providing additional context for the output of the inpainting module to be more consistent to the ground truth. Fig. 2 shows the AME module, which extracts a mask corresponding to a test image. Considering that liver tumors usually have density lower than that of normal tissues, an adaptive thresholding strategy is employed to exclude the higher-density tissues, which is regarded as normal tissues in this context.

In this study, we focus on the liver region, initially segmenting it using a liver mask, which can be easily extracted using current segmentation algorithms [48, 49]. After extracting the liver region, we unfold the image into a vector with the background outside the liver being removed. We select the median value as the threshold, which we call the adaptive threshold because it varies case by case. As expected, the tumor should have a lower density than the threshold. We further binarize the difference between the adaptively thresholded liver region and the original liver region at another empirical threshold, which was set to 20HU in this study, to form the mask.

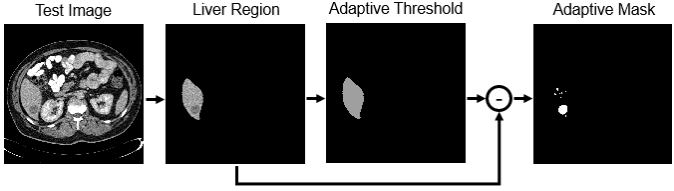

Figure 2. Pipeline of adaptive mask extraction. The liver mask is used to extract the liver region for each test image. The adaptive threshold is obtained by calculating the median value in the liver region. The mask is constructed by binarizing the different map between the liver region and adaptive threshold images.

### C. Image Inpainting

The goal of inpainting is to predict masked pixels in an image. Denote the test image as $x$, the unknown pixels as $m \odot x$ and the known pixels as $(I - m) \odot x$. Since every reverse step from $x_t$ to $x_{t-1}$ depends solely on $x_t$, altering the known regions $(I - m) \odot x_t$ is permissible if the correct properties of the corresponding distribution are maintained. Given that the forward process is defined by a Markov chain of added Gaussian noise, sampling the intermediate image $x_t$ at any point in time is possible. This allows sampling the know regions $m \odot x_t$ at any time step $t$. Thus, in terms of the known region and unknown regions, one reverse step of the approach is expressed as

$$x_{t-1}^{known} \sim \mathcal{N}\big(\sqrt{\bar{\alpha}_t}x_0, (1 - \bar{\alpha}_t)I\big) \quad (12)$$

$$x_{t-1}^{unknown} \sim \mathcal{N}\big(\sqrt{\bar{\alpha}_t}x_0, (1 - \bar{\alpha}_t)I\big) \quad (13)$$

$$x_{t-1} = m \odot x_{t-1}^{known} + (1 - m) \odot x_{t-1}^{unknown} \quad (14)$$

where $x_{t-1}^{known}$ is sampled using the known pixels in the given image $m \odot x_0$, while $x_{t-1}^{unknown}$ is obtained from the model, given the previous iteration $x_t$. They are then combined into the new sample $x_{t-1}$ using the mask, as ilusrated in Fig. 3. The afornmentioned mask has an erea generally smaller than the liver region. Some images similar to the image of interest can be found in the memery bank, and used as a prompt to facilitate more accruacy inpainting.

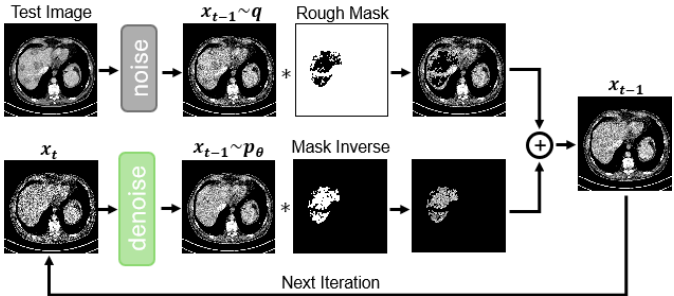

Figure 3. Overview of the inpainting module. In each step, the known region is extracted from the input, and the masked region is inpainted by DDPM.

However, the combination of $x_{t-1}^{known}$ and $x_{t-1}^{unknown}$ may introduce inconsistencies. Since the DDPM was trained to generate images within the original data distribution, it naturally aims to produce consistent structures. This DDPM property is used to harmonize the model output. Specifically, the output $x_{t-1}$ is fiffused back to $x_t$ by sampling as $x_t \sim \mathcal{N}\big(x_t; \sqrt{1 - \beta_t}x_{t-1}, \beta_t I\big)$, which is a resampling step. As a result, some information incorporated in the generated region $x_{t-1}^{unknown}$ is still preserved in $x_t^{unknown}$. This leads to a new $x_t^{unknown}$ that is more harmonized with $x_t^{known}$ and contains inpainted information. Applying the resampling operation $r$ times incorporates the desired semantic information over the entire liver region [43].

### D. Anomaly Score

The generative diffusion prior is employed to obtain the inpainted image, $x_{repaint}$. The anomaly score between the test image and inpainted image is defined as

$$score(i,j) = F\big((x_{repaint}(i,j) - x_{test}(i,j))\big) \quad (15)$$

where $F(\cdot)$ is the median filter used to suppress the noise. If this anomaly score is greater than the threshold σ, it is judged as an anomaly; otherwise, it is considered as normal. The optimal σ is the threshold value that achieves the maximal F1-score during the calculation of the area under the curve (AUC). That is, test images will have binary prediction results, classified as either anomalous or normal. Since only the liver region is inpainted, the anomaly scores focus on detecting liver tumors. Finally, interpretable heatmaps are obtained to estimate tumor areas, assisting radiologists in their reporting.

## III. EXPERIMENTS

### A. Datasets

Two datasets, including the liver tumor segmentation benchmark (LiTS) [51] and the image reconstruction for comparison of algorithm database (IRCAD) [52], were employed to evaluate our proposed method. The LiTS dataset was used for training and testing, while the IRCAD dataset was only used to test the proposed method.

**LiTS:** The LiTS dataset was organized in conjunction with the IEEE International Symposium on Biomedical Imaging (ISBI) 2017 and the International Conferences on Medical Image Computing and Computer-Assisted Intervention (MICCAI) 2017 and 2018. The data and segmentations were provided by several clinical sites worldwide. This study consists of 130 abdominal CT scans from the challenge as the dataset for anomaly detection. Each scan includes tumors with liver segmentation masks and liver tumor annotations. We selected 11,035 healthy slices from the 130 CT scans, which do not include any tumors, as the dataset for training the DDPM. We selected 716 abnormal slices, each containing at least one tumor, as the test dataset. The test cohort covers diverse types of liver tumor diseases, including hepatocellular carcinoma, cholangiocarcinoma, metastases from primary colorectal, breast and lung cancers. The images were acquired with different CT scanners and acquisition protocols, including various image noise.

**IRCAD:** The IRCAD database consists of CT scans of 10 women and 10 men, with 15 cases including hepatic tumors. Each scan includes tumors with liver segmentation masks and liver tumor annotations. The dataset features major difficulties the liver segmentation software may encounter due to contact with neighboring organs, atypical liver shape and density, and/or image noise. We selected 113 abnormal slices, each containing at least one tumor, as the test dataset to assess the proposed method.

### B. Implementation Details

Our DDPM approach was implemented in the PyTorch 2.2.2 framework, trained and evaluated on an H100 GPU. For training the model, we normalized the pixel values to the range [-1, 1] using $4u - 1$, where $u$ is the attenuation coefficient in the CT images. We used the Adam optimizer with a learning rate of $10^{-4}$. A linear sequence was employed for the variance schedule, with the start and ending values of betas set to $10^{-4}$ and 0.02, respectively. We tracked an exponential moving average (EMA) of the model parameters with a momentum of 0.999. The batch size was set to 1. The training was conducted in 1,500,000 iterations over one week. For inpainting, we used $T = 250$ timesteps and $r = 10$ times. The EMA parameters were employed during inpainting.

### C. Evaluation Metrics

A collection of relevant metrics was used to evaluate the methods used in this study. Some metrics discussed here were derived from the four basic cardinalities of the confusion matrix namely true positive (TP), true negative (TN), false positive (FP) and false-negative (FN).

**The receiver operator characteristic (ROC) curve (AUC):** The receiver operator characteristic (ROC) curve is a plot that visualizes the tradeoff between the true positive rate (TPR) and the false-positive rate (FPR) at various threshold values. The area under the ROC curve (AUC) is the measure of the ability of a classifier to distinguish between classes and is used as a summary of the ROC curve.

$$TPR = \frac{TP}{TP + FN} \quad (16)$$

$$FPR = \frac{FP}{FP + TN} \quad (17)$$

**Average precision (AP):** Average precision (AP) summarizes the precision recall (PR) curve to one scalar value, where a PR curve plots the value of precision against recall for different confidence threshold values. AP is high when both precision and recall are high, and low when either of them is low across a range of confidence threshold values. The range for AP is between 0 to 1.

$$Precision = \frac{TP}{TP + FP} \quad (18)$$

$$Recall = \frac{TP}{TP + FN} \quad (19)$$

**DICE:** The DICE score evaluates the degree of overlap between the binarized anomaly scores and reference annotations. It is also called an overlapping index. In addition, to compare the segmentation result with ground truth data, DICE also measures reproducibility. The value of DICE lies in the interval [0, 1] where 1 is the perfect detection. Given two binary masks A and B, it is formulated as:

$$Dice(A, B) = \frac{2|A \cap B|}{|A| + |B|} \quad (20)$$

### D. Competing Methods

We compare our method with several competing methods, including:

**Autoencoders (AE):** AE can effectively compress and represent normal variations while simultaneously restricting its generalization ability, rendering it incapable of representing abnormal changes. This makes AE widely used for anomaly detection and serves as a baseline in this study.

**Variational Autoencoders (VAE):** There are several variants of AE, and we select VAE as a representation method.

**Denoising Autoencoders (DAE):** While AE and its variants may inadvertently leak abnormal information to the decoder, resulting in deteriorated performance, DAE uses a denoising training approach to harness powerful reconstruction capability. The DAE method has demonstrated the best performance among twenty-seven anomaly detection methods in the BraTS2021 dataset [20], making it the state-of-the-art method selected for this study.

We use the code from the paper by Cai et al. [20], which conducts a comparative study of anomaly detection in medical images. All parameters were set according to their recommendations.

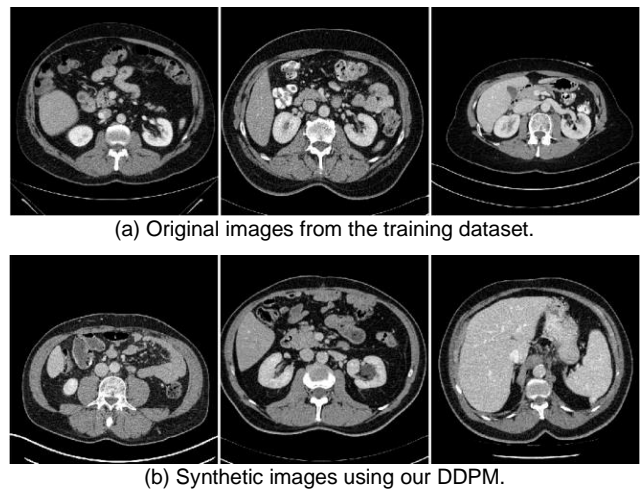

(a) Original images from the training dataset.

(b) Synthetic images using our DDPM.

Figure 4. Six representative liver CT images in the original and synthetic datasets respectively. The display window is [-125, 225] HU.

## IV. RESULTS

### A. Synthesis Data

We trained the improved DDPM on the LiTS training dataset. The synthetic images using the DDPM are shown in Fig. 4. Quite different from natural images, CT images are usually influenced by image noise limited by x-ray radiation. In Fig. 4(a), we can observe that the original images from the training dataset contain noise, and the noise level may be influenced by different factors such as tube current and patient body size. Also, the liver region has arbitrary shapes, and CT values also vary among different images. In Fig. 4(b), synthetic images show high image quality. Note that the synthesized images have realistic structural details and noise patterns.

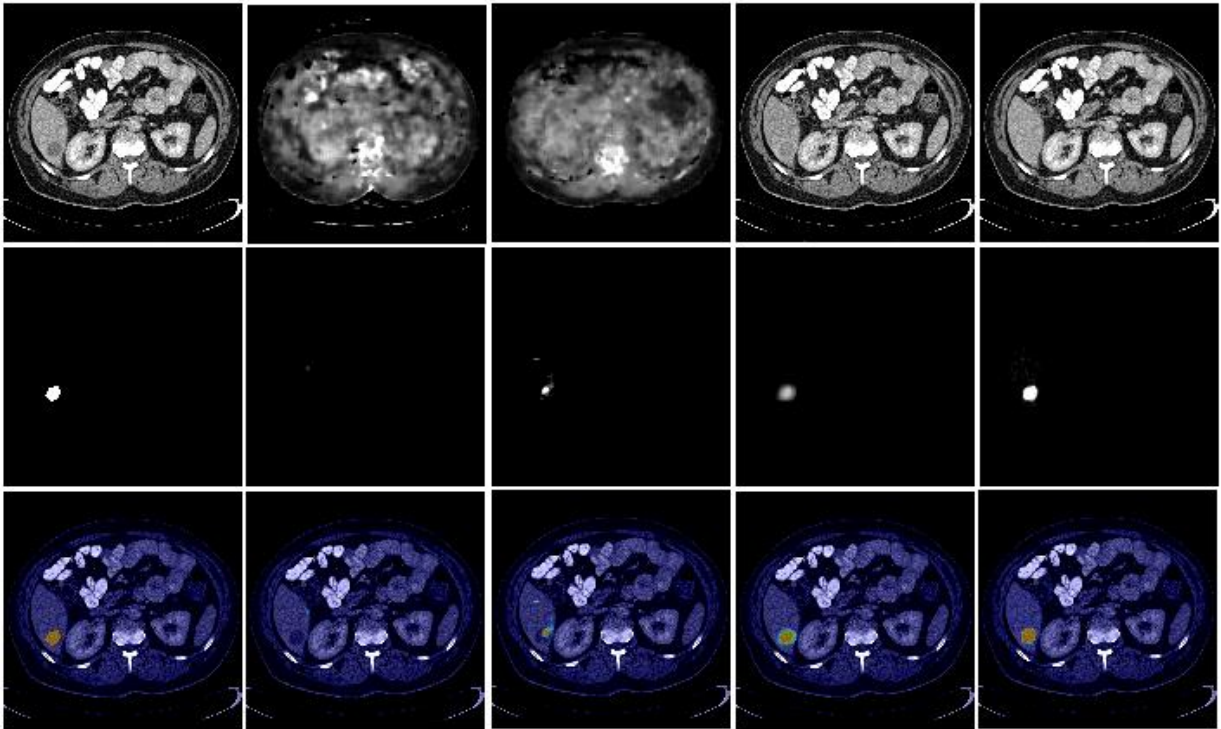

Figure 6. Visualization of the compared anomaly detection methods on an image with only one tumor from the LiTS dataset. From left to right are the input images, the image reconstructed by AE, VAE, DAE and GDPI. The first row displays the input abnormal or reconstructed images in a display window of [-125, 225] HU. The second row shows the reference annotation of the tumor along with the anomaly images in a display window of [0, 50] HU. The last row presents the heatmaps of the input abnormal images or reconstructed images. The color bar ranging from blue to orange corresponding to 0 to 50 HU.

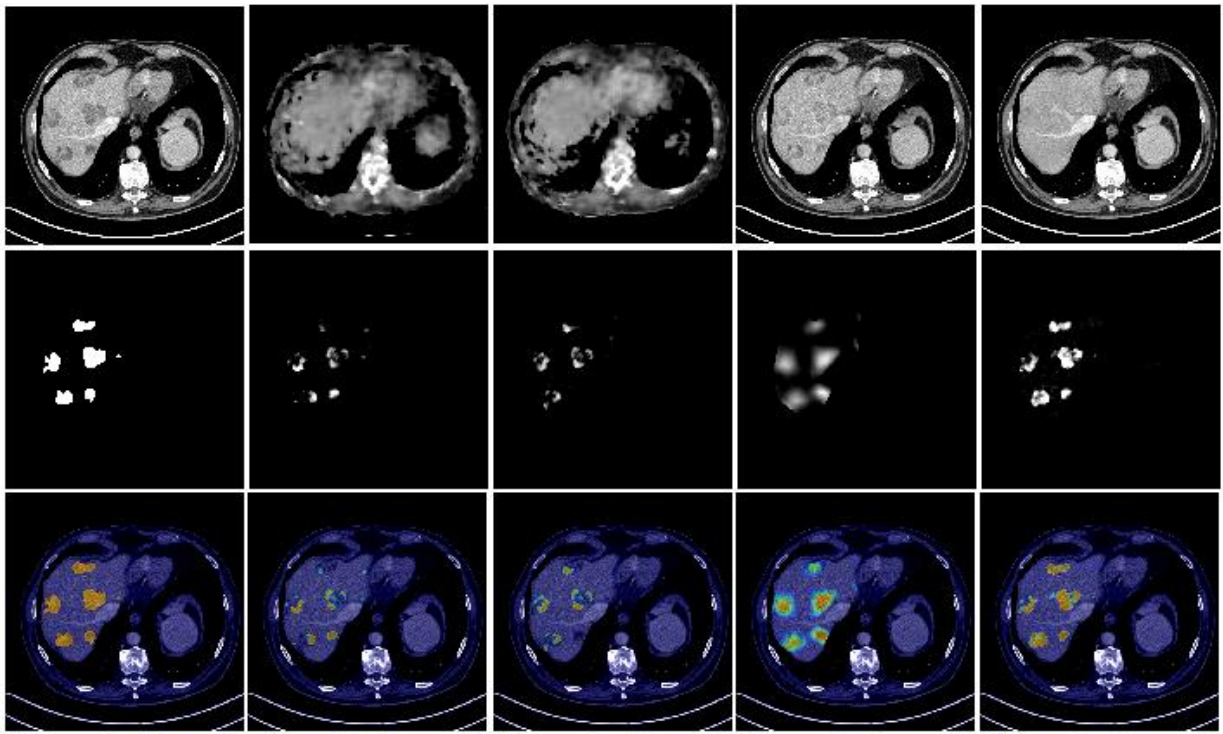

Figure 7. Visualization of the compared anomaly detection methods on an image with multiple tumors from the LiTS dataset, with the same arrangement of the images as that in Fig. 6.

## B. Results on The LiTS Dataset

We first evaluated the performance on the LiTS test dataset. Fig. 5 shows the ROC curves for the methods compared in this study. The VAE method achieved the lowest performance while the AE method performed better than VAE, though the results were still comparable. The DAE method significantly improved performance. In contrast, our proposed method achieved the best performance in terms of ROC. For quantitative evaluation, Table I presents the metrics for different methods, following a trend similar to that indicated by the ROC results. The AE and VAE methods did not perform well, while the DAE method showed tremendous improvement in DICE, AUC, and AP scores. Our proposed method achieved the best performance across all quantitative measures on the entire test dataset.

For visualization, we compared anomaly detection methods using two representative slices: one with a single tumor and one with multiple tumors. Fig. 6 illustrates the reconstructed images in a single tumor scenario. Both the AE and VAE methods reconstructed distorted images with blurred structures, especially outside the liver region. Since the liver region does not include complex structures, the anomaly score for AE did not show major differences, with only one small point in the normal region incorrectly identified as a liver tumor. The AE method failed to detect the liver tumor, partially due to the low contrast in CT images. The VAE method detected part of the liver tumor but still had false alarms in normal regions. The DAE method avoided errors in normal regions and successfully detected and localized the liver tumor. However, the liver tumor remained in the reconstructed image, leading to a lower anomaly score amplitude. The heatmap in the tumor region was also blurred compared to the reference. Our proposed method reconstructed the best image visually matched to the normal image, with the tumor region inpainted by normal tissue. The anomaly score matched well with the reference annotation, as confirmed by the heatmap.

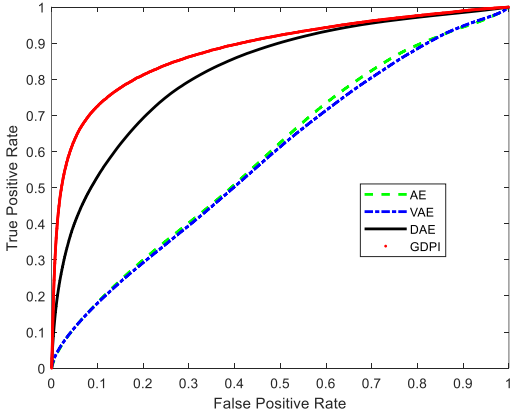

Figure 5. ROC curves of different methods.

TABLE I: ANOMALY DETECTION PERFORMANCE METRICS OF DIFFERENT ALGORITHMS ON THE LiTS DATASET.

|  | AUC | AP | DICE |
|---|---|---|---|
| AE | 59.34 | 15.45 | 21.84 |
| VAE | 58.55 | 15.47 | 21.32 |
| DAE | 81.67 | 43.76 | 44.68 |
| GDPI | **88.14** | **62.24** | **62.05** |

Fig. 7 shows results for multiple tumors. This case included six tumors: five large ones and one extremely small tumor. The AE method localized all tumors, but the shapes of the anomaly scores differ significantly from the reference, and there was one detection error in a normal region. The VAE method performed similarly to AE but failed to detect one tumor. The DAE method detected the five large tumors with a blurred anomaly score, missed the small tumor, and had one detection error in a normal region. Our proposed method detected all tumors. However, the area of the small tumor was smaller than the reference, possibly due to the denoising module. The anomaly scores for the five large tumors also differ from the reference, due to the unsupervised nature of our approach. While the anomaly score amplitude was more consistent with the grayscale value of the tumor than the expert's annotation, there is still room to improve the detectability. These results highlight the potential of our method to assist radiologists in identifying and localizing liver tumors.

### C. Robustness Test on The IRCAD Dataset

**Abnormal Liver Tumor Detection:** Fig. 8 shows the results for an image with multiple tumors from the IRCAD dataset. In this case, two tumors are included: a large tumor and a small one. The AE method localized both the tumors but included one detection error in a normal region. The VAE method mistakenly detected several tumors in normal tissues. The DAE method detected the tumors with a blurred anomaly score, misclassified a normal tissue piece between the two tumors as a tumor and had one detection error in a normal region. Our proposed method reported all the tumors. However, the area of the small tumor is smaller than the reference, possibly influenced by the denoising module and the fact that the radiologist provided only a tumor mask without specifying the tumor density.

Table II shows the quantitative results for the IRCAD dataset. The AE and VAE methods perform better than they did on the LiTS dataset. This is likely because both AE and VAE reconstruct distorted images, making them insensitive to differences between datasets. In contrast, the DAE and GDPI methods show decreased performance on the IRCAD dataset due to the distribution variation between the two datasets and the coarse tumor annotations provided for IRCAD. Despite these observations, our proposed GDPI method achieved the best performance on the IRCAD dataset, demonstrating the robustness of the proposed method.

TABLE II: ANOMALY DETECTION PERFORMANCE METRICS OF DIFFERENT ALGORITHMS ON THE IRCAD DATASET.

|  | AUC | AP | DICE |
|---|---|---|---|
| AE | 62.80 | 29.71 | 34.29 |
| VAE | 65.62 | 31.95 | 37.81 |
| DAE | 65.51 | 41.97 | 43.03 |
| GDPI | **76.36** | **57.29** | **58.49** |

**Normal Image Bias:** Table III presents a comparative analysis of bias in normal images across different methods. Since the normal images do not include tumors, the previously mentioned metrics are not applicable in this scenario. To assess normal image bias across various methods, we calculated the mean value of the score map. A smaller mean value indicates less bias between the reconstructed image and the normal reference image. Although the GDPI method achieves the smallest mean value, some bias remains between the reconstructed and reference images, as low-density normal tissues, such as pneumobilia, may still be detected as tumors. This phenomenon will be discussed in detail in the discussion section.

TABLE III: COMPARATIVE ANALYSIS OF BIAS IN NORMAL IMAGES.

|  | AE | VAE | DAE | GDPI |
|---|---|---|---|---|
| Mean | 1.0812 | 3.5575 | 3.4580 | 1.0564 |

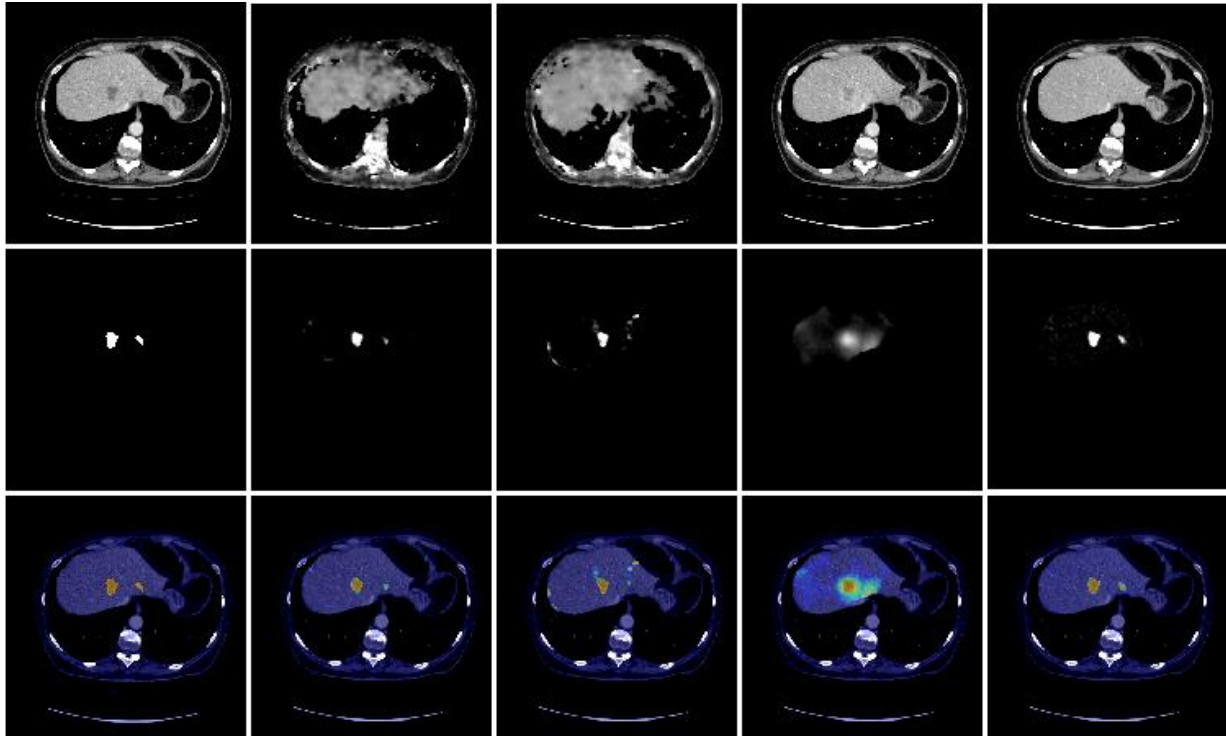

Figure 8. Visualization of the compared anomaly detection methods on an image with multiple tumors from the IRCAD dataset, with the same arrangement of the images as that in Fig. 6.

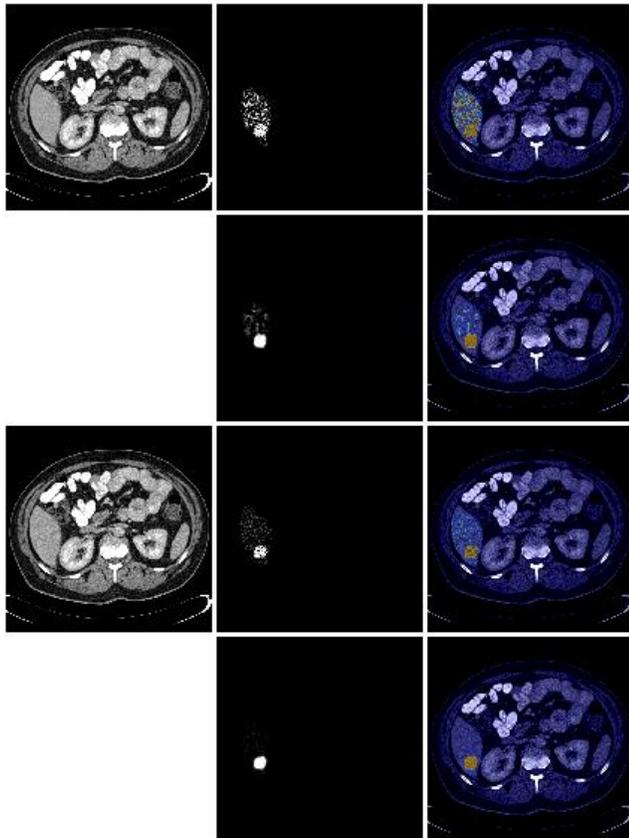

Figure 9. Visualization of the ablation study findings. The first column displays the reconstructed images. The second column is the anomaly scores. The third column is the heatmaps. From top to bottom are the images reconstructed by GDPI without masking and denoising, GDPI without masking, GDPI without denoising and GDPI, respectively.

### D. Ablation Study

The proposed GDPI method consists of the four main modules: DDPM, AME, Inpainting, and Anomaly scoring modules. To evaluate the influence of each module, we conducted the four ablation studies: 1) GDPI-WRD: This method does not use either the AME nor denoising modules; 2) GDPI-WD: This method uses the AME module but not the denoising module; 3) GDPI-WR: This method uses the denoising module but not the AME module; 4) GDPI: This is the proposed method, using both the AME and denoising modules. The reconstructed results are shown in Fig. 9. It is observed that the anomaly score of the GDPI-WRD method is influenced by severe noise, leading to detection errors in some normal regions. The GDPI-WD method effectively reduces noise in the anomaly score, but there are still detection errors in some normal regions. The GDPI-WR method can eliminate coarse-grained structural errors, but the anomaly score is still affected by noise. The anomaly score of our proposed method closely matches the reference annotation, demonstrating its effectiveness in this context. The quantitative results in Table IV further demonstrate the effectiveness of the proposed method.

TABLE IV: ANOMALY DETECTION PERFORMANCE METRICS KEYED TO EACH OF THE NETWORK CONFIGURATIONS.

|  | AUC | AP | DICE |
| --- | --- | --- | --- |
| GDPI-WRD | 65.09 | 16.97 | 27.14 |
| GDPI-WR | 70.32 | 21.78 | 32.74 |
| GDPI-WD | 74.01 | 29.81 | 36.96 |
| GDPI | **88.14** | **62.24** | **62.05** |

## V. DISCUSSIONS

Unlike natural images, CT images are of lower contrast, especially between normal liver tissue and tumors. The AE and its variant VAE fail to detect liver tumors because they reconstruct distorted images where key information is lost. Both DAE and GDPI methods can reconstruct high-quality images, leading to more accurate tumor detection. Since CT images usually include noise, in this study we employed a median filter to suppress noise. The results demonstrate that while noise can be suppressed to some extent, the anomaly score map becomes blurry, and some tiny tumors may be eliminated. Using more advanced denoising methods may enhance the performance of the proposed method.

The shape of the liver is quite variable. Directly using the inpainting method to inpaint the whole liver region may introduce mismatches between the inpainted image and the original abnormal image. Additionally, structures outside the liver region may be inpainted into the liver region because no specific conditions are imposed for the liver generation. In our future work, we will propose a shape-guided and class-guided diffusion model to train the model to prevent the inpainting module from inpainting anatomical structures outside the liver region.

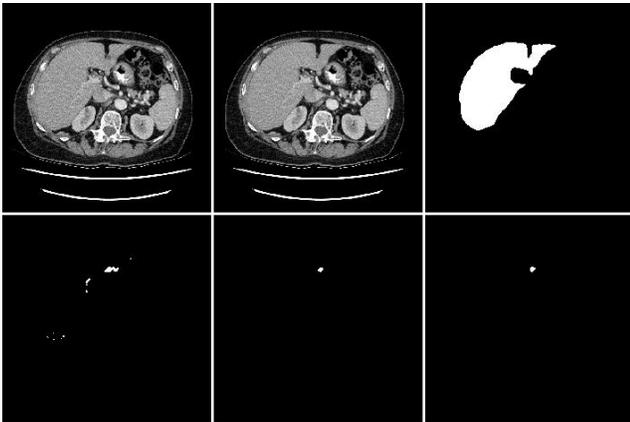

Figure 10. Exemplary of an adaptive mask. In the first row, from left to right, are the input image, the reconstructed image, and the liver mask. In the second row, from left to right, are the adaptive mask, the binarized anomaly score and the reference annotation.

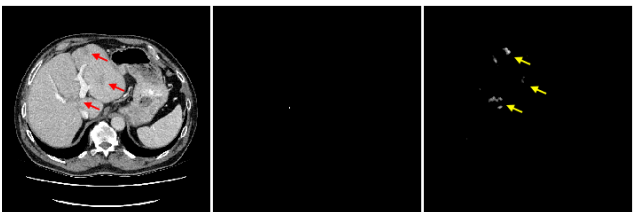

Figure 11. Exemplary detection errors. From left to right are the input abnormal image, the reference annotation, and the anomaly score.

The CT values of tumor and healthy liver tissues vary due to different scanner settings. We use an inpainting module to reconstruct the abnormal regions. Since our inpainting module is trained on a dataset with various scanner settings, and the healthy liver tissue excluded by the adaptive mask provides a prompt, the inpainted regions closely resemble the surrounding healthy tissues. Consequently, we obtain the anomaly score by calculating the difference between the inpainted healthy tissues and the abnormal tumors, which remains consistent across different scanner settings. However, once we separate the residuals from the whole images, we will further analyze the high-order statistics, such as texture and deep features, to enhance the anomaly score.

We empirically selected an adaptive threshold to get an adaptive mask based on the knowledge that liver tumors usually have lower density than normal liver tissues. Fig. 10 shows an example of an adaptive mask. We can observe that the CT values of most healthy tissues exceed the threshold value, causing them to be excluded by the adaptive mask. However, there are some areas extracted by the adaptive mask that are not labeled as tumors in the reference annotation. Using the adaptive mask, the results can be further enhanced by our proposed GDPI method, leading to a more accurate binarized anomaly score compared to the reference annotation. In addition, there are some low-density areas that resemble tumors but are not labeled as such by radiologists. For example, the regions indicated by arrows in Fig. 11. The radiologist only labeled one tiny point as the liver tumor, but there are multiple low-density regions that may be benign tumors. In addition, some low-density normal tissues, such as pneumobilia, may be detected as tumors, which particularly affects performance in normal images. To enhance the tumor detection performance in this scenario, in our future work we will further cascade a classification module after the proposed GDPI to distinguish among healthy tissues, benign and malignant liver tumors. Training on additional sets that includes a range of benign and malignant liver findings will also be necessary.

## VI. CONCLUSION

In this paper, we have proposed a generative diffusion prior-based inpainting method for CT liver tumor anomaly analysis. We have demonstrated that the proposed method achieves a 7.9% improvement in AUC on the LiTS dataset. The robustness of our approach has been validated on another dataset. We believe that our method can be further improved and potentially incorporated into radiology CAD systems for liver tumor detection and localization.